\documentclass[12pt]{iopart}

\usepackage{graphicx}
\begin{document}

\title[URC dark matter density profile for Pulsar in M31 and M87 Galaxies ]{Possible Existence of Dark Matter Admixed Pulsar in M31 and M87 Galaxies}

\author{Sulagna Mondal$^1$, Sajahan Molla$^2$, Prabir Kumar Halder$^3$ and Mehedi Kalam$^4$}

\address{$^1$ Department of Physics, Amity University, Kolkata, India. \\
	$^2$ Department of Physics, New Alipore College, L Block, New Alipore, Kolkata 700053, West Bengal, India. \\
	$^3$ Department of Physics, Coochbehar Panchanan Barma University, Coochbehar, West Bengal, India. \\ 
	$^4$ Department of Physics, Aliah University, IIA/27, New Town, Kolkata 700160, West Bengal, India.  }
\ead{sulagna.mondal21@gmail.com, sajahan.phy@gmail.com, prabirkrhaldar@gmail.com, kalam@associates.iucaa.in(corresponding author)}

\vspace{10pt}
\begin{indented}
\item[]Keywords: compact star; dark matter; mass function; radius; compactness; red-shift
\end{indented}

\begin{abstract}
In previous studies~\cite{EurPhysJPlus.135.362,EurPhysJPlus.135.637,Universe.8.652} the possible existence dark matter admixed pulsar have been discussed based on three different dark matter density profiles, Singular Isothermal Sphere density profile, Universal Rotational Curve density profile and Navarro-Frenk-White density profile. They have been used to discuss the pulsars present in our Milky Way galaxy as well as some satellite dwarf galaxies of Milky Way. In this article we use the Universal Rotational Curve (URC) dark matter density profile to observe similar effects on galaxies M31 and M87. These study hold significant importance, as now it can be concluded that there is a fair possibility of presence of dark matter admixed pulser in M31 and M87 galaxies as well.
\end{abstract}

%
%
%
%
%

\section{Introduction}

Compact objects like neutron stars, white dwarfs, strange stars are getting more and more attention from the astrophysicists owing to their unique properties and development of new observation techniques. These objects can act as excellent stellar laboratories to study various astrophysical principles as well as to find out new lesser-known characteristics of the universe. Compact objects are mainly made up of Fermi gas, the outward degeneracy pressure produced from this is balanced by the inward gravitational force. This Fermi gas, in case of white dwarfs is made up of electrons, whereas in neutron stars they are mainly neutrons. Strange quark stars on the other hand are constituted of strange quark matters present at the core of neutron stars \cite{PhysRevD.89.043014,Astron.Astrophys.160.121}. The strong interaction presents among these quarks, along with the gravitational attraction make such strange stars extra stable and more bound than normal neutron stars. The formation of such strange stars requires huge amount of energy which is accounted for by the massive energy released during the occurrence of a super luminous supernovae \cite{Mon.Not.R.Astron.Soc.387.1193}. A strange star can be differentiated from an ordinary neutron star on the basis of their vanishing surface energy density \cite{Astron.Astrophys.160.121,Astrophys.J.310.261,PhysRevD.30.2379,PhysRevD.82.024016,Phys.Lett.B.438.123}. After the birth of a neutron star, its temperature falls rapidly and reaches bellow the Fermi energy, according to a review work done by Lattimer and Prakash \cite{Phys.Rep.442.1.109}, for a given equation of state (EoS), the mass and radius of a neutron star depend only on central density. In case of a spherically symmetric compact star their mass and radius can be estimated theoretically from the solutions of Tolman-Oppenheimer-Volkoff (TOV) equations. They can also be measured by pulsar timing, thermal emission from cooing stars, surface explosions and gravitational wave emission through observations. It is a challenging task to fix an EoS \cite{PhysRevD.80.103003,Astrophys.J.693.2.1775,PhysRevD.82.101301,Astrophys.J.712.2.964,Astrophys.J.719.2.1807} due to the complex structure of the compact stars. Although from observational data mass has been calculated for a number of such stars present in binaries \cite{Mon.Not.R.Astron.Soc.305.132,Astron.Astrophys.303.497,Astrophys.J.387.340,PhysRevLett.94.111101,Mon.Not.R.Astron.Soc.286.L21}, there is very little information about their radius. Therefore, theoretical study of these objects is important to support the observational findings. In this context the following studies could be of interest \cite{Gen.Relativ.Gravit.44.107,Eur.Phys.J.C.72.2017,Mod.Phys.Lett.A.31.40,Mod.Phys.Lett.A.32.4,Astrophys.Space.Sci.362.10.188,Int.J.Mod.Phys.D.21.1250088,Class.Quantum.Grav.23.1525,PhysRevLett.96.251101,Eur.Phys.J.C.76.5.266,Astrophys.Space.Sci.361.160D,Eur.Phys.J.Plus.129.3,Astrophys.Space.Sci.357.74,Astrophys.Space.Sci.356.2.327,Int.J.Theor.Phys.53.11.3958,Eur.Phys.J.C.77.9,Res.Astron.Astrophys.19.026,Astrophys.SpaceSci.364.112,JCAP.07.004,AstrophysJ.848.24,Eur.Phys.J.Plus.135.819}.

In 1933 Swiss astronomer Fritz Zwicky first introduced the concept of dark matter while studying the speed of Coma cluster \cite{Phys.Acta.6.110,Gen.Relativ.Gravit.41.207}. Later, the optical studies of galaxies (e.g., M31) by Rubbin and Ford \cite{Astrophys.J.159.379} presented the same conclusion about dark matter. There are various concepts regarding the origin and nature of dark matter, and some of them could well explain its properties. Although ordinary matters have very little to no direct interaction with dark matter, they have some remarkable gravitational effects on stellar bodies \cite{Astropart.Phys.37.70,J.Cosmol.Astropart.Phys.10.031,PhysRevD.96.023002,Eur.Phys.J.C.75.2.41}. It has been observed that fermionic dark matter particles have a significant influence on the physical properties of the strange stars \cite{PhysRevD.96.083013,PhysRevD.74.063003,PhysRevD.85.103528,PhysRevD.93.083009,PhysRevD.96.083004}. The proposed idea of self-interaction among dark matter particles by Spergel and Steinherdt \cite{PhysRevLett.84.3760} could potentially resolve some of the conflicts between non-interacting cold dark matter and observed phenomena.

Some astrophysicists have previously worked on the topic of dark matter in neutron stars \cite{PhysRevD.84.107301,PhysLettB.725.200,PhysRevD.87.123506,Astrophys.J.835.1.33,Int.J.Mod.Phys.D.27.16.1950002}. Inspired by their works scientists have studied the existence of dark matter along with ordinary baryonic matter within neutron stars. It has been investigated by using the Singular Isothermal Sphere dark matter density profile\cite{EurPhysJPlus.135.362}. Here it has considered that dark matter is mixed with ordinary matter with the following density distribution \cite{astro.ph.0102341}: $\rho_{d} (r)=\frac{K}{2\pi G r^2}$ , 
where K is the velocity dispersion. It has been shown the possible existence of dark matter along with ordinary matter within well-known pulsars namely PSR J1748-2021B in NGC 6440B, PSR J1911-5958A in NGC 6752, PSR B1802-07 in NGC 6539, and PSR J1750-37A in NGC 6441 galaxies. We also investigated the existence of dark matter in pulsars by using the Navarro-Frenk-White (NFW) density profile particularly in the disk regions of the Milky Way galaxy \cite{Universe.8.652}. 

Now we choose the Universal Rotational Curve dark matter density profile since it is a more accepted model and is applicable throughout the central as well as the outer region of the galactic halo. The URC dark matter density profile is therefore given as \cite{Natural Science 4.265,Astrophysical Journal.447.1.L25},
\begin{equation}
	                                \rho_{d} (r)=  \frac{\rho_{0} r_{0}^3}{(r+r_0)(r^2+r_{0}^2)}                                 
\end{equation}
Where $r_0$ is the core radius and $ \rho_{0}$ is the central density. The values of the parameters $r_0$ and $ \rho_{0}$ are different for different galaxies. Previously it has shown the existence of some well-known pulsars in the Milky Way galaxy \cite{EurPhysJPlus.135.637} (in which case the values were $r_0=9.11 kpc$ and $\rho_{0}=5 \times 10^{-24} \times (\frac{r_{0}}{8.6 kpc}) ^{-1} $ gm/c.c. respectively \cite{Natural Science 4.265,Astrophysical Journal Letters.744.1.L9}). In this study, using the URC dark matter density profile we will discuss about the possibility of existence of dark matter admixed pulsar in other galaxies such as Andromeda galaxy(M31) \cite{Astrophysical Journal.685.1.254} ($r_0=21 $ kpc and $\rho_{0}$=0.011 $M_{\odot} pc^{-3}$) and M87 \cite{Physical Review D.100.044012} ($r_0=91.2$kpc and $\rho_{0}=6.9 $$\times$ $10^{6}$ $M_{\odot} / kpc^{3}$) respectively.

Therefore, our plan of work is as follows: In Sect.2, we discuss about the interior spacetime of a pulsar. In Sect.3, we look into some physical properties of pulsars in M31. We discuss the behaviour of energy density and pressure, matching condition with exterior Schwarzschild solution, compactness, surface redshift, energy condition and validity of generalized TOV equation for M31 galaxy. In Sect.4, We study our result for the same on a  X-ray pulsar 3XMM J004301.4+413017 \cite{Astrophysical Journal.839.125}. In Sect.5, we discuss the same physical properties stated above for M87 galaxy. However, due to absence of the data of any detected known pulsar in the M87, in Sect.6, we have considered a model pulsar of an average mass to justify our result, if in future any such pulsar is detected in M87, it would be interesting to study their properties in light of our proposed model. In Sect.~7 we put our discussion and concluding remarks.

\section{Interior Spacetime}
\par The interior spacetime associate with the spherically symmetric pulsar \cite{Eur.Phys.J.C.79.8.706,Eur.Phys.J.C.78.6.437,Eur.Phys.J.C.75.8.389,PhysRevD.99.044029,Eur.Phys.J.A.53.6.141,Eur.Phys.J.A.53.5.89,Eur.Phys.J.A.54.11.207,Chin.Phys.C.41.015103,Mod.Phys.Lett.A.32.08.1750053,Astrophys.Space.Sci.361.11.352,Astrophys.Space.Sci.361.262,Astrophys.Space.Sci.361.5.163,Pramana.89.2.23} is described by the line element as
\begin{equation}\label{eq:1}
	ds^{2}=-e^{\nu(r)}dt^{2}+e^{\lambda(r)}dr^{2}+r^{2} d\theta^{2}+r^{2} \sin\theta^{2} d\phi^{2}
\end{equation}
\par The energy-momentum tensor for the matter distribution in the interior of the isotropic pulsar is assumed to
be
\begin{equation}\label{eq:2}
	T_{\nu}^{\mu}=diag (-\rho,p,p,p)
\end{equation}
where $\rho$ and $p$  are the energy density, pressure respectively.
The Einstein's field equations which reflect the fact that the matter distribution affect the topology of the spacetime are written in
geometric unit $(G=c=1)$ as
\begin{eqnarray}\label{eq:3}
	8 \pi \rho  &=e^{-\lambda}\left(\frac{\lambda'}{r}-\frac{1}{r^{2}}\right)+\frac{1}{r^{2}} \label{eq:3a} \\
	8 \pi p
	&=e^{-\lambda}\left(\frac{\nu'}{r}+\frac{1}{r^{2}}\right)-\frac{1}{r^{2}}
	\label{eq:3b} \\
	8\pi p &=
	\frac{e^{-\lambda}}{2}\left(\frac{(\nu^\prime)^2 - \lambda^\prime
		\nu^\prime}{2} + \frac{\nu^\prime - \lambda^\prime}{r} +
	\nu^{\prime\prime}\right) \label{eq:3c}
\end{eqnarray}
\par We are considering Heintzmann metric \cite{Z.Phys.228.489}. According to Heintzmann
\begin{eqnarray}\label{eq:4}
	e^{\nu} &=A^{2} (1+ a r^{2})^{3} \label{eq:4a}\\
	e^{-\lambda} &=1-\frac{3 a r^{2}}{2}\left[\frac{1+C(1+4 a r^{2})^{-\frac{1}{2}}}{1+a r^{2}}\right] \label{eq:4b}
\end{eqnarray}
where $A$ and $C$ are dimensionless constant and $a$ is a constant with dimension of $length^{-2}$ in geometric unit.

\par  Now, we consider the pulsars are made of ordinary matter admixed
with condensed dark matter. Therefore, effective density and
pressure can be written as
\begin{eqnarray*}
	\rho_{eff} &=\rho + \rho_{d}  \\
	p_{eff}   &= p  - p_{d}
\end{eqnarray*}
where pressure due to dark matter, $p_d = m \rho_d$, m is a
constant \cite{MNRAS.449.1.403}.

\par Now, Einstein's field equations for the metric ansatz (2) and eqn.(1) are modified in presence of dark matter as
\begin{eqnarray}\label{eq:5}
	\rho  &=\frac{1}{8 \pi}\left[e^{-\lambda}\left(\frac{\lambda'}{r}-\frac{1}{r^{2}}\right)+\frac{1}{r^{2}}\right]-
	\frac{\rho_{0} r_{0}^{3}}{(r+r_{0})(r^{2}+r_{0}^{2})}  \label{eq:5a}\\
	p  &=\frac{1}{8 \pi}\left[e^{-\lambda}\left(\frac{\nu'}{r}+\frac{1}{r^{2}}\right)-\frac{1}{r^{2}}\right]+\frac{m \rho_{0} r_{0}^{3}}{(r+r_{0})(r^{2}+r_{0}^{2})}\label{eq:5b}
\end{eqnarray}

\section{Study of Physical Properties for Pulsar in Galaxy M31}
\subsection{Energy Density and Pressure}
From the plot of energy density and pressure with radius (see fig.1 and fig.2), we see that both energy density and pressure are maximum at the center and decreases monotonically towards the boundary. Thus, both pressure and density are well behaved in the interior of the stellar structure.
Here we have set the values of the constants to be $a = 0.004 km^{-2}, C = 0.6, m = 0.055$, $r_0=21 kpc$ and $\rho_{0}$=0.011 $M_{\odot} pc^{-3}$ such that all required conditions must be obeyed, including pressure, which reduces to zero at the boundary.
 
\begin{figure}
	\centering
	\includegraphics[width=0.7\linewidth]{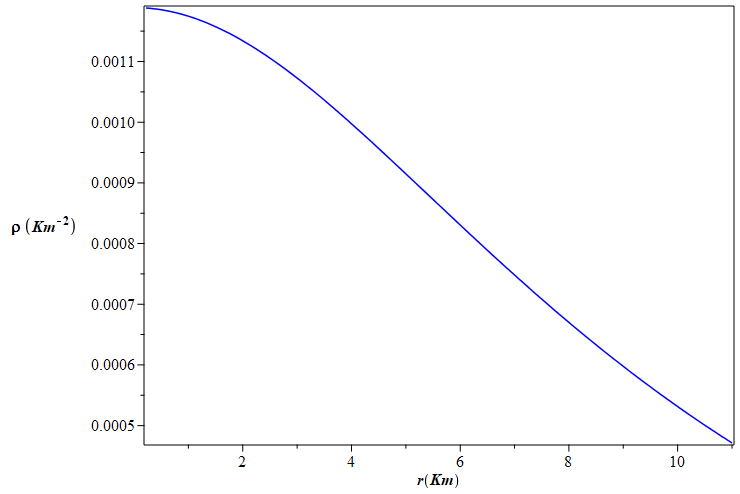}
	\caption{Energy density versus radius for $a = 0.004 km^{-2}, C = 0.6, m = 0.055$, $r_0=21 kpc$ and $\rho_{0}$=0.011 $M_{\odot} pc^{-3}$}
	\label{fig:energy-conditions-m31}
\end{figure}
\begin{figure}
	\centering
	\includegraphics[width=0.7\linewidth]{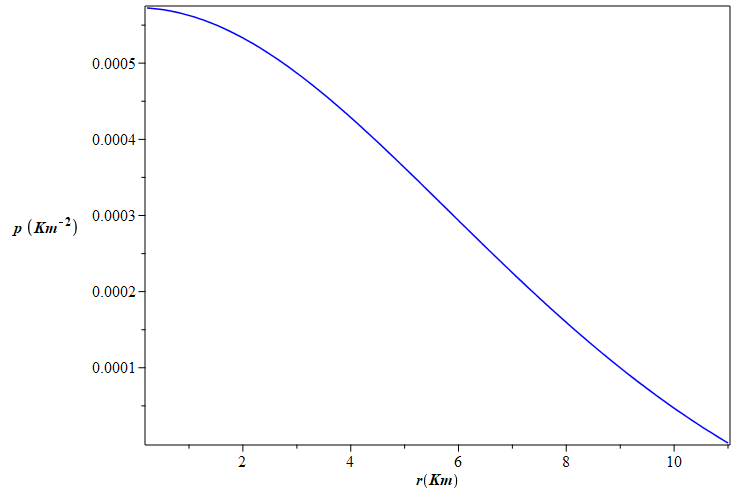}
	\caption{Pressure versus radius for $a = 0.004 km^{-2}, C = 0.6, m = 0.055$, $r_0=21 kpc$ and $\rho_{0}$=0.011 $M_{\odot} pc^{-3}$}
	\label{fig:p-m31}
\end{figure}

\subsection{Energy Conditions}
A realistic material must satisfy all the energy conditions namely, Null energy condition (NEC), Weak energy condition (WEC), Strong energy condition (SEC) and Dominant energy condition (DEC) at all points in the interior of the stellar object or compact star. These conditions are represented as follows:\\
(i) NEC: $\rho \geq 0$                        \\
(ii) WEC: $\rho+p  \geq 0$, $p \geq 0$         \\
(iii) SEC: $\rho+p  \geq 0$, $\rho+3p  \geq 0$ \\
(iv) DEC: $\rho >|p| $ \\
\begin{figure}
	\centering
	\includegraphics[width=0.7\linewidth]{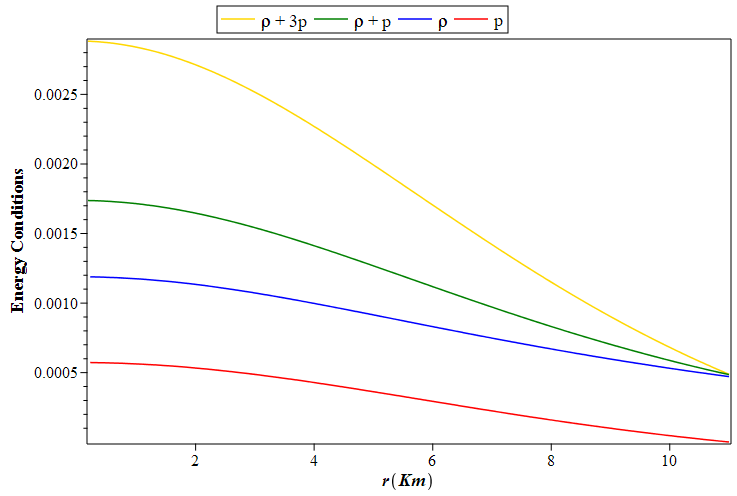}
	\caption{Energy conditions versus radius for $a = 0.004 km^{-2}, C = 0.6, m = 0.055$, $r_0=21 kpc$ and $\rho_{0}$=0.011 $M_{\odot} pc^{-3}$}
	\label{fig:energy-conditions-m31}
\end{figure}
Fig.~3 shows that all these energy conditions are satisfied in our proposed stellar model.

\subsection{Matching conditions}
At the boundary (r = R) the interior metric of a stellar body must match with the exterior Schwarzschild metric,
\begin{equation}\label{eq:7}
	ds^{2}=-(1-\frac{2 M}{r})dt^{2}+\frac{dr^{2}}{(1-\frac{2 M}{r})}+r^{2}(d\theta^{2}+\sin\theta^{2} d\phi^{2})
\end{equation}
Assuming continuity of the metric function $g_{tt}$, $g_{rr}$ and $\frac{dg_{tt}}{dr}$ across the surface of the star $(r=R)$ we obtain,
\begin{eqnarray}
	e^{\nu(R)} &=1-\frac{2 M}{R} \label{eq:8a}\\
	e^{-\lambda(R)} &=1-\frac{2 M}{R} \label{eq:8b}\\
	\nu' e^{\nu(R)} &=\frac{2 M}{R^{2}}\label{eq:8c}
\end{eqnarray}
From this the gravitational mass of the pulsar is found to be,
\begin{equation}
	M =\frac{3 a R^{3}}{4}\left[\frac{1+C(1+4 a
		R^{2})^{-\frac{1}{2}}}{1+a R^{2}}\right]
\end{equation}

\subsection{Mass-Radius relation and Surface red-shift}
The calculation of the radial dependence of gravitational mass function m(r) yields,
\begin{equation}
	m(r)=4\pi\int^{r}_{0} \rho_{eff}~~ \tilde{r}^2 d\tilde{r} = \frac{3ar^3\left[1+C(1+4ar^2)^{-\frac{1}{2}}\right]}{4\left(1+ar^2\right)}
\end{equation}
Therefore, the compactness of the star is,
\begin{equation}
	u(r)=\frac{m(r)}{r}=\frac{3ar^2 \left[ 1+C\left(1+4 a
		r^{2}\right)^{-\frac{1}{2}}\right]}{4\left(1+a r^{2}\right)}
\end{equation}
The corresponding surface redshift of the star can be written as,
\begin{equation}
	Z_{s}=\frac{1}{\sqrt{1-2 u}}-1=\frac{1}{\sqrt{1-\frac{3 a r^2
				\left(\frac{C}{\sqrt{4 a r^2+1}}+1\right)}{2 \left(a
				r^2+1\right)}}}-1
\end{equation}
\begin{figure}
	\centering
	\includegraphics[width=0.7\linewidth]{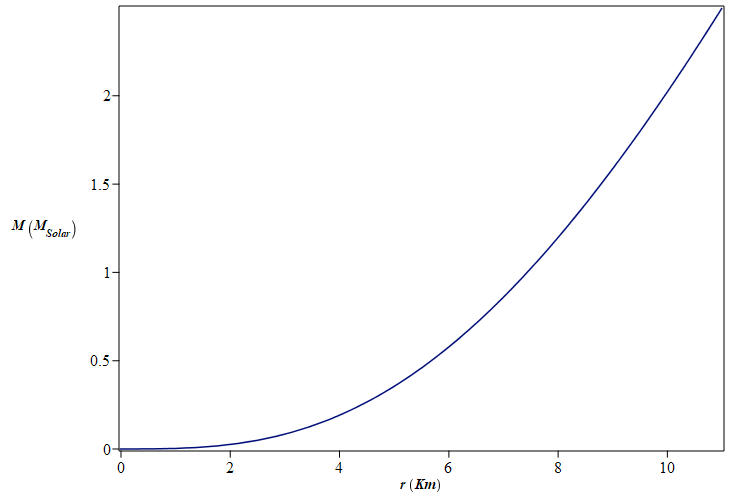}
	\caption{Radial dependance of mass function for $a = 0.004 km^{-2}, C = 0.6, m = 0.055$, $r_0=21 kpc$ and $\rho_{0}$=0.011 $M_{\odot} pc^{-3}$}
	\label{fig:m-m31}
\end{figure}
\begin{figure}
	\centering
	\includegraphics[width=0.7\linewidth]{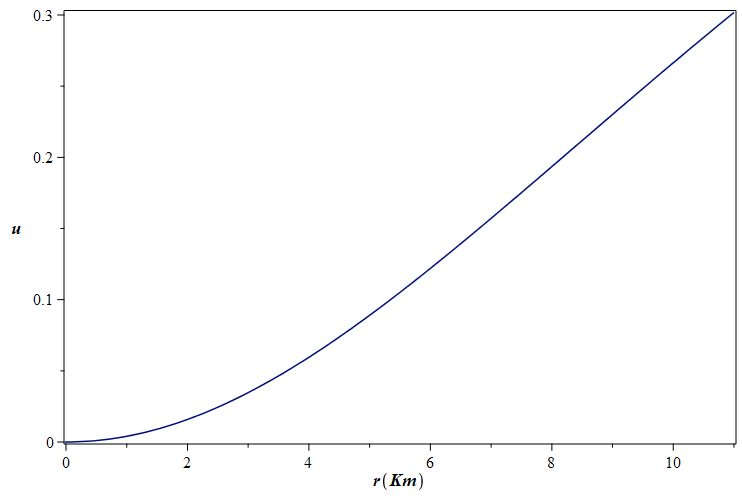}
	\caption{Compactness versus radius for $a = 0.004 km^{-2}, C = 0.6, m = 0.055$, $r_0=21 kpc$ and $\rho_{0}$=0.011 $M_{\odot} pc^{-3}$}
	\label{fig:u-m31}
\end{figure}
\begin{figure}
	\centering
	\includegraphics[width=0.7\linewidth]{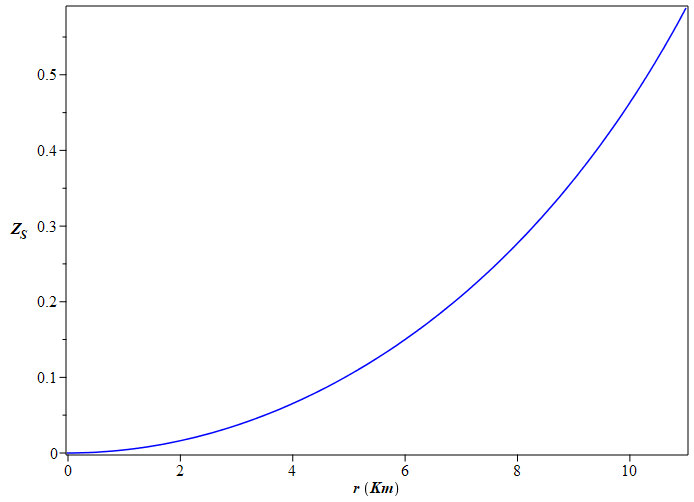}
	\caption{Redshift versus radius for $a = 0.004 km^{-2}, C = 0.6, m = 0.055$, $r_0=21 kpc$ and $\rho_{0}$=0.011 $M_{\odot} pc^{-3}$}
	\label{fig:z-m31}
\end{figure}
The Mass function of the pulsar is shown in fig.~4. It is evident from Fig.5 that the compactness, $u(r)$ is an increasing function with the radius. According to the Buchdahl condition \cite{PhysRev.116.1027} of compactness, the maximum value of u(r) should not exceed $ \frac{4}{9}$ =0.444, in our case the maximum value of compactness i.e.,$ u(r)$ is found to be 0.31. Likewise, Fig.6. shows that, the maximum value of surface redshift in our case is $Z_s$ = 0.584, which is much less than the allowed maximum surface redshift value $(Z_s\leq0.85)$ \cite{Nucl.Phys.Proc.Suppl.80.1110}. Hence our model satisfies these two conditions as well \cite{Astronomy and Astrophysics.530.A137}.

\subsection{Generalized TOV Equation}
For an isotropic system, the generalized TOV equation can be written as
\begin{equation}
	\frac{dp}{dr}+\frac{1}{2} \nu' (\rho+p)=0
\end{equation}
The stable equilibrium condition for the stellar body under the effect of gravitational force $F_{g}$ and hydrostatic force $F_{h}$ is represented by the following equation,
\begin{equation}
	F_{g}+F_{h}=0
\end{equation}
Here
\begin{eqnarray}
	F_{g} &=-\frac{1}{2} \nu' (\rho+p)\\
	F_{h} &=-\frac{dp}{dr}
\end{eqnarray}
\begin{figure}
	\centering
	\includegraphics[width=0.7\linewidth]{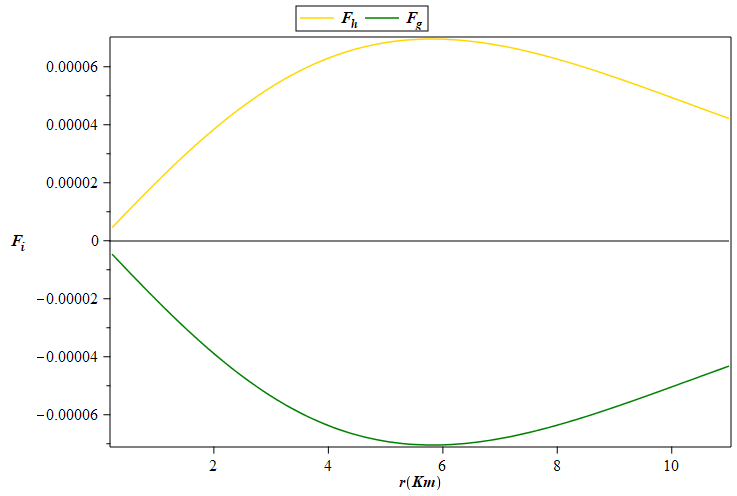}
	\caption{Gravitational force $F_{g}$ and hydrostatic force $F_{h}$ versus radius for $a = 0.004 km^{-2}, C = 0.6, m = 0.055$, $r_0=21 kpc$ and $\rho_{0}$=0.011 $M_{\odot} pc^{-3}$}
	\label{fig:tovm31}
\end{figure}

In Fig.7 we can see that the gravitational force and hydrostatic force are in perfect equilibrium in the interior region of the compact star.

\section{Measurement of the physical properties}
From the fig.~8 we measure possible radius of the pulsar and thereafter we calculate the other physical properties (see table \ref{tab:1}) of the X-ray pulsar J004301.4+413017
\begin{figure}
	\centering
	\includegraphics[width=0.7\linewidth]{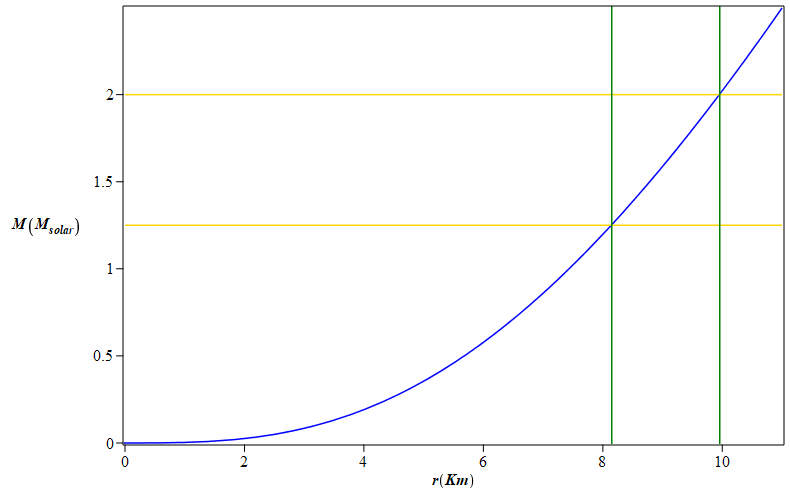}
	\caption{Possible radii of the X-ray pulsar J004301.4+413017}
	\label{fig:mass-radius-m31}
\end{figure}

\begin{table*}
	\caption{Evaluated parameters for pulsars}
	\label{tab:1}       
\begin{tabular}{|c|c|c|c|c|}
	\hline
Pulsar	& Mass($M_{\odot}$) & Radius(km) from model & Compactness & Redshift \\
	\hline
	3XMM J004301.4+413017 & 1.25 (min)& 8.14 & 0.198 & 0.289 \\
	\hline
	3XMM J004301.4+413017 & 2.0 (max)& 9.95 & 0.265 & 0.460 \\
	\hline
\end{tabular}
\end{table*}

\section{Study of physical properties for Pulsar in the Galaxy M87}
\subsection{Energy density and pressure}
In case of M87, from the plot of energy density(see fig.~9) and pressure(see fig.~10) with radius, we see that both energy density and pressure are maximum at the center and decreases monotonically towards the boundary. Thus, both pressure and density are well behaved in the interior of the stellar structure.\\
Here we have set the values of the constants to be a = 0.003 $km^{-2}$, C = 0.76, m = 0.09, $r_0=91.2$kpc and $\rho_{0}=6.9 $$\times$ $10^{6}$ $M_{\odot} / kpc^{3}$ $=5 \times 10^{-24} \times (\frac{r_{0}}{8.6 kpc}) ^{-1} $ gm/c.c. such that all required conditions must be obeyed, including pressure, which reduces to zero at the boundary.
\begin{figure}
	\centering
	\includegraphics[width=0.7\linewidth]{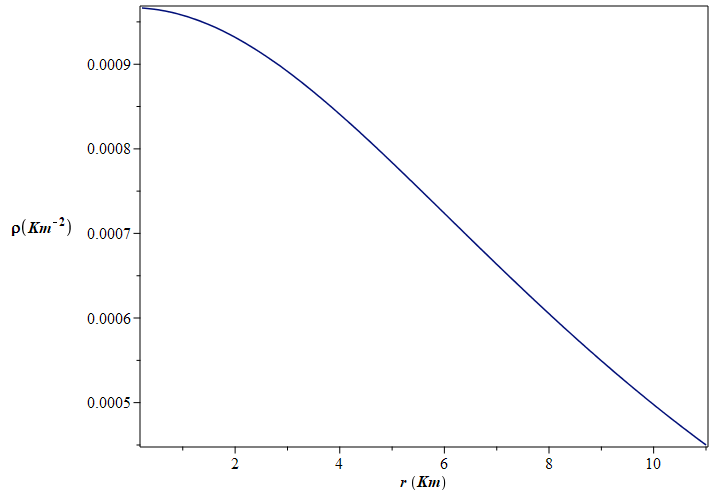}
	\caption{Energy density versus radius for a = 0.003 $km^{-2}$, C = 0.76, m = 0.09, $r_0=91.2$kpc and $\rho_{0}=6.9 $$\times$ $10^{6}$ $M_{\odot} / kpc^{3}$ }
	\label{fig:rho-m87}
\end{figure}
\begin{figure}
	\centering
	\includegraphics[width=0.7\linewidth]{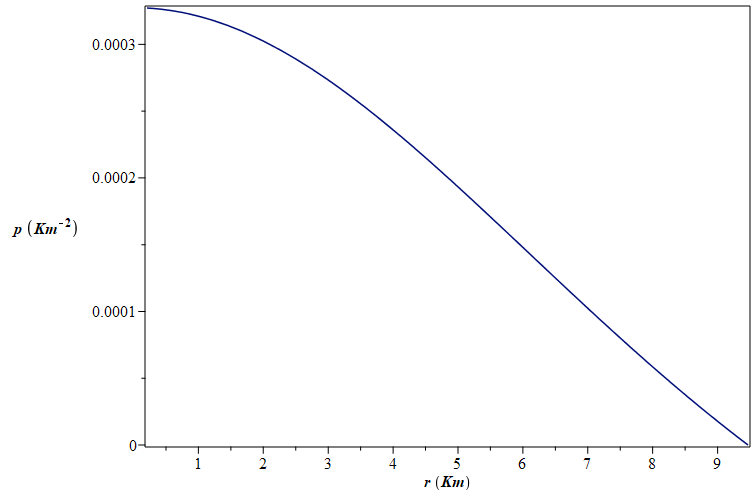}
	\caption{Pressure versus radius for a = 0.003 $km^{-2}$, C = 0.76, m = 0.09, $r_0=91.2$kpc and $\rho_{0}=6.9 $$\times$ $10^{6}$ $M_{\odot} / kpc^{3}$ }
	\label{fig:p-m87}
\end{figure}
The other conditions and equations for M87 are the same as M31, hence we do not repeat the same calculations, rather we now enclose the plots of physical properties for M87 with its specified set of parameters which we have already discussed above.

\subsection{Energy conditions}
The four energy conditions are satisfied in case of pulsars present in M87. The energy conditions are as follows:\\
(i) NEC: $\rho \geq 0$                        \\
(ii) WEC: $\rho+p  \geq 0$, $p \geq 0$         \\
(iii) SEC: $\rho+p  \geq 0$, $\rho+3p  \geq 0$ \\
(iv) DEC: $\rho >|p| $ \\
\begin{figure}
	\centering
	\includegraphics[width=0.7\linewidth]{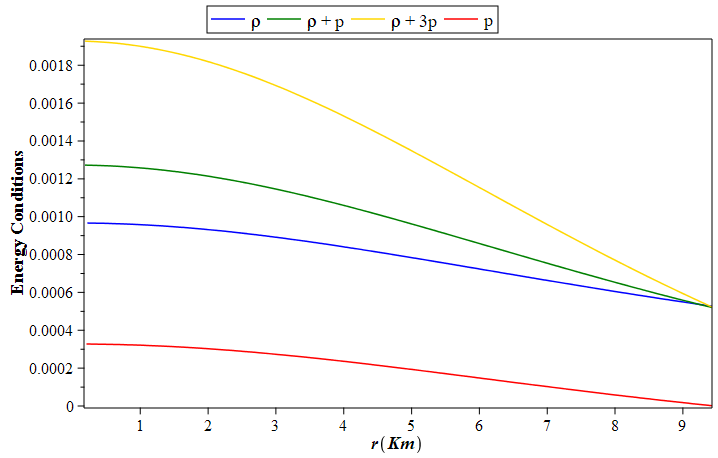}
	\caption{Energy conditions versus radius for a = 0.003 $km^{-2}$, C = 0.76, m = 0.09, $r_0=91.2$kpc and $\rho_{0}=6.9 $$\times$ $10^{6}$ $M_{\odot} / kpc^{3}$ }
	\label{fig:energy-conditions-m87}
\end{figure}
Fig.11. shows that they behave in our predicted manner.

\subsection{Mass-Radius relation and Surface Redshift}
Like before, matching Schwarzschild condition at the boundary of the star gives us eqns. (12), (13), (14). From there we can find the gravitational mass M (see eqn. (15)) of the stellar object.\\
In Fig.s 12, 13 and 14 we see that mass function, compactness and surface redshift of the compact star behaves well while varying with the radius, while considering the new set of parameters for M87 galaxy.
\begin{figure}
	\centering
	\includegraphics[width=0.7\linewidth]{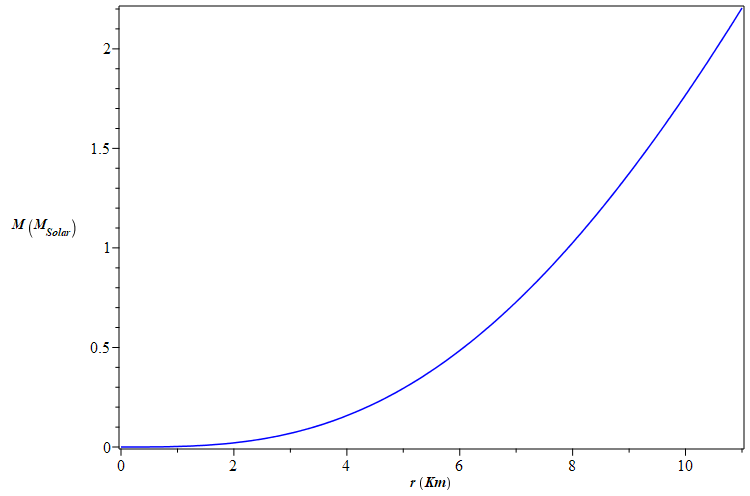}
	\caption{Mass function versus radius for a = 0.003 $km^{-2}$, C = 0.76, m = 0.09, $r_0=91.2$kpc and $\rho_{0}=6.9 $$\times$ $10^{6}$ $M_{\odot} / kpc^{3}$ }
	\label{fig:m-m87}
\end{figure}
\begin{figure}
	\centering
	\includegraphics[width=0.7\linewidth]{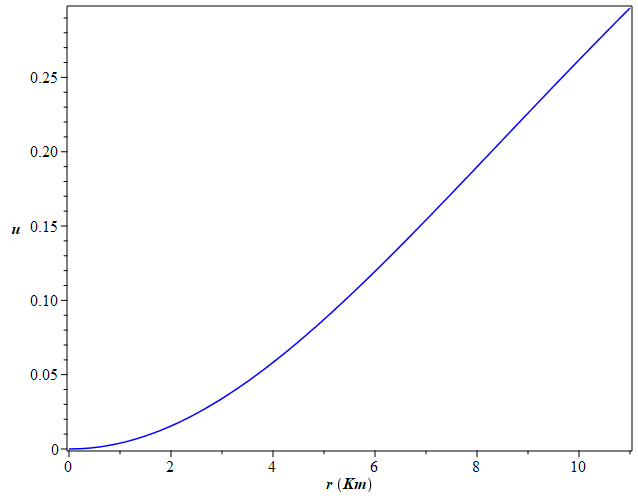}
	\caption{Compactness versus radius for a = 0.003 $km^{-2}$, C = 0.76, m = 0.09, $r_0=91.2$kpc and $\rho_{0}=6.9 $$\times$ $10^{6}$ $M_{\odot} / kpc^{3}$ }
	\label{fig:u-m87}
\end{figure}
\begin{figure}
	\centering
	\includegraphics[width=0.7\linewidth]{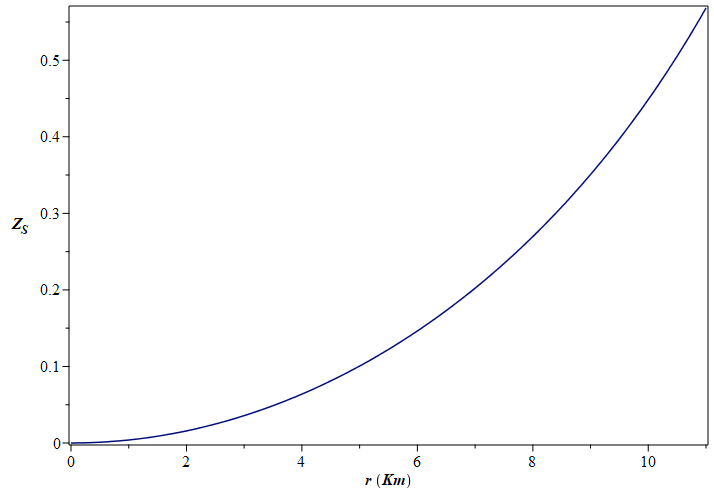}
	\caption{Redshift versus radius for a = 0.003 $km^{-2}$, C = 0.76, m = 0.09, $r_0=91.2$kpc and $\rho_{0}=6.9 $$\times$ $10^{6}$ $M_{\odot} / kpc^{3}$ }
	\label{fig:z-m87}
\end{figure}
We see in Fig.13 that the compactness, $u(r)=\frac{m(r)}{r}$ is an increasing function with the radius, it’s highest value being u(r) = 0.293 which satisfies the Buchdahl condition \cite{PhysRev.116.1027} of mass-radius ratio. Likewise, Fig.14. shows that, the maximum value of surface redshift in this case is $Z_s$ = 0.582, which is much less than the allowed maximum surface redshift value $(Z_s\leq0.85)$. Hence our model satisfies the required conditions for pulsars residing in the M87 galaxy. 

\subsection{Generalized TOV Equation}
The generalized TOV equation (eq.~19) is satisfied within the interior of the structure as the gravitational and hydrostatic forces are in perfect equilibrium (eq.~20). 
\begin{figure}
	\centering
	\includegraphics[width=0.7\linewidth]{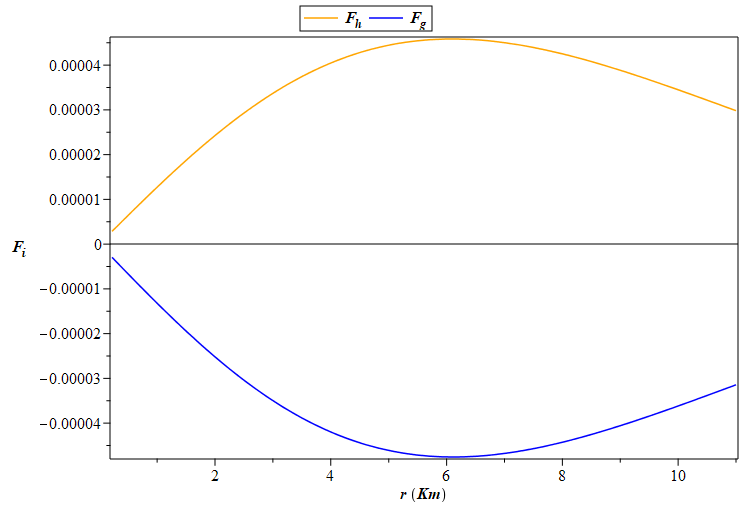}
	\caption{Gravitational force ($F_g$) and hydrostatics force ($F_h$ ) versus radius for a = 0.003 $km^{-2}$, C = 0.76, m = 0.09, $r_0=91.2$kpc and $\rho_{0}=6.9 $$\times$ $10^{6}$ $M_{\odot} / kpc^{3}$ }
	\label{fig:tov-m87}
\end{figure}

From the expressions of gravitational force and hydrostatic forces (eq.~21 and eq.~22) we can see that they maintain well equilibrium (Fig.~15), this contributing to the stability of the stellar structure.

\section{Model Pulsar's physical parameter measurement}
As we can see from the previous figures, our model successfully satisfies all the conditions of pulsar properties. Due to lack of available data being reported for a pulsar in M87 it is not possible to justify this model by any observed data for a pulsar in M87. Hence, we take here a model pulsar named “Model PSR M87” to show that my proposed model satisfies its properties as well in Galaxy M87.
We are considering a model pulsar “Model PSR M87” having mass 1.65$M_{\odot}$ and we calculate its radius (see Fig.16), compactness and redshift (see table~\ref*{tab:2}).

\begin{figure}
	\centering
	\includegraphics[width=0.7\linewidth]{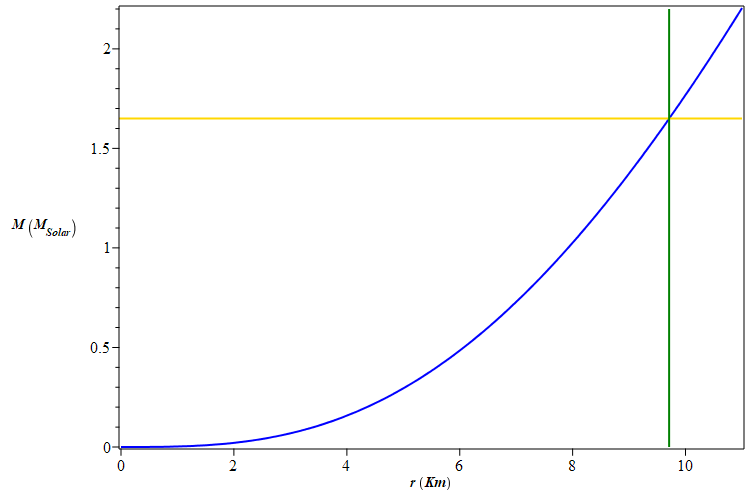}
	\caption{Possible radii of the pulsar "Model PSR M87"}
	\label{fig:mass-radius-m87}
\end{figure}
\begin{table*}
	\caption{Evaluated parameters for pulsar in Galaxy M87}
	\label{tab:2}       
\begin{tabular}{|c|c|c|c|c|}
	\hline
	Pulsar & Mass($M_{\odot}$) & Radius (in km)  & Compactness  & Redshift  \\
	\hline
	Model PSR M87 & 1.65 & 9.71 & 0.251 & 0.419 \\
	\hline
\end{tabular}
\end{table*}

\section{Discussion and concluding remarks }
In the previous studies scientists have considered a two-fluid model where a pulsar is formed of ordinary matter admixed with dark matter. For this they have taken up three different dark matter density profiles, the Singular Isothermal Sphere (SIS) profile, Universal Rotational Curve (URC) or the Burkert Profile and Neverro-Frank-White (NFW) profile. Of these three the SIS model is the simplest one, the NFW profile fits well to the disk region of the galaxy whereas the URC profile is applicable throughout the halo of the galaxy. Hence, these previous studies encourage us to investigate if we can use the URC density profile and apply this same two-fluid dark matter admixed pulsar model, to pulsars which reside at outer galaxies other than our Milky Way. 

In this article, we have taken core radii ($r_{0}$) and central densities ($\rho_{0}$) of two different galaxies M31, also known as the Andromeda galaxy and M87 in the Virgo cluster, to carry out our investigation. With the use of the URC density profile ($\rho_{d} (r)=  \frac{\rho_{0} r_{0}^3}{(r+r_0)(r^2+r_{0}^2)} $) we have considered the values of constants in the metric ($a = 0.004 km^{-2}, C = 0.6, m = 0.055$, $r_0=21 kpc$ and $\rho_{0}$=0.011 $M_{\odot} pc^{-3}$ for M31 and a = 0.003 $km^{-2}$, C = 0.76, m = 0.09, $r_0=91.2$kpc and $\rho_{0}=6.9 $$\times$ $10^{6}$ $M_{\odot} / kpc^{3}$) in such a way that the pressure vanishes at the boundary. We have successfully shown that our proposed model can satisfy all the necessary physical properties such as, density and pressure in the interior of the pulsar are well behaved, they both are maximum at the central region and decreases monotonically towards the boundary. 

We have also seen that our pulsar model satisfies all the energy conditions, the generalized TOV equation. From the mass function we can find out all the important features at the interior of a pulsar, it further satisfies the Buchdahl condition of compactness ($\frac{2M}{R}<8/9$). The highest value of Redshift as found in our model is much less than the maximum allowed value ($Z_{s}\leq0.85$). From the mass-radius relationship we have found out the radius of the pulsar 3XMM J004301.4 +413017, the slowest spinning X-ray pulsar in M31 galaxy. 

Although our model satisfies all the physical properties in both the galaxies, due to lack of any actual pulsar data reported in the M87 till date, we could not fit any observational data. However, we have considered a model pulsar of mass 1.65$M_{\odot}$ and shown that it justifies the proposed model. This study henceforth opens the possibility of further studies on actual pulsar data in M87 if they happen to be detected in the near future.

Thus, as a concluding remark it can be said that this study is rather important as it successfully solidifies the possibility of existence of dark matter (using the URC profile) admixed with ordinary matter within pulsars, not only in the Milky Way galaxy, but also in other giant galaxies such as M31 and M87. This is perfectly in alignment with the previous studies \cite{EurPhysJPlus.135.362,EurPhysJPlus.135.637,Universe.8.652} where we predicted the possibility of such pulsars in most of the galaxies. We therefore highly encourage to carry out further study related to this at other galaxies where a remarkable number of pulsars has already been detected.

\section*{Acknowledgments}
SM and MK would like to thank IUCAA, Pune, India for providing research facilities and warm hospitality under Visiting Associateship where a part of this work was carried out.PKH is also thank to IUCAA, Pune, India for providing research facilities and warm hospitality where a part of this work was carried out. 

\section*{Data Availability Statement}
This manuscript has no associated data or the data will not be deposited. [Authors' comment: This manuscript has no measured data associated; the plots involve data generated by modelling.]

\section*{References\label{except}}

\end{document}